\documentclass[reprint,aps,prl,amsmath,amssymb]{revtex4-1}

\usepackage{graphicx}
\usepackage{dcolumn}
\usepackage{bm}

\begin{document}

\title{Boundary induced amplification and nonlinear instability of interchange modes}

\author{Jupiter Bagaipo}
\email{jbagaipo@umd.edu}
\author{A. B. Hassam}
\address{Institute for Research in Electronics and Applied Physics, University of Maryland, College Park, MD 20742-3511}

\date{\today}

\begin{abstract}
It is shown that small distortions on the boundaries are amplified in the core of a magnetized plasma if the system is close to marginal stability for the ideal magnetohydrodynamic interchange mode. It is also shown that such marginal systems can be nonlinearly unstable. The combination of boundary amplification and nonlinearity is shown to result in a nonlinear instability. The induced instability is highly sensitive to the boundary in that, if the fractional deviation from marginality is a small parameter $b$, the system can go unstable from fractional boundary distortions of $\mathcal{O}(b^{3/2})$.
\end{abstract}

\maketitle

\emph{I. Introduction.}---Magnetically confined plasmas for fusion are limited in how much pressure can be contained, by the so-called $\beta$ limit -- the critical ratio of pressure to magnetic energy density.\cite{Freidberg-book} This limitation generally comes from interchange instabilities, wherein flux tubes of high pressure plasma can interchange with outer, lower pressure flux tubes.\cite{Freidberg-book, Kulsrud-stellarator, Longmire} Such energy release can be stabilized if the flux tube interchange is disallowed by topology (on account of the frozen-in condition for strongly magnetized plasmas), that is to say, if the ``transverse'' field is strong enough.\cite{Longmire} For maximum efficiency, one wants to operate close to marginal stability, $\beta \rightarrow \beta_c$.

In this letter, we report on two findings pertaining to operating near marginal stability. Let $\Delta\beta=\beta_c-\beta$ and let the system size be $a$. Then, (1) we establish the general idea that a small perturbation of $\delta/a$ on the boundary \emph{amplifies} interchange displacements in the core of the plasma, by an amplification factor $\beta_c/\Delta\beta$; (2) we show that the system is \emph{nonlinearly unstable}, so that a critical boundary perturbation will destabilize the interchange mode even for systems below the linear $\beta$ stability limit. Upon combining these two findings, we find that the amplification phenomenon leads to a nonlinear instability criterion which is highly sensitive to boundary perturbations, namely the fractional critical size is even smaller than the fractional deviation from marginality, i.e., $(\delta/a) > |\Delta\beta/\beta_c|^{3/2}$. This has the implication that magnetic configurations designed to confine plasma close to the $\beta$ limit within a tolerance of $\epsilon$ would necessitate that the design be more sensitive to boundary perturbations; specifically, boundary tolerances need to be better than $\epsilon^{3/2}$.  Such considerations are of significant importance in the design of axisymmetric tolerances for advanced tokamaks as well as in the fully 3D design of stellarators for fusion.

Field amplification near marginal stability has been shown to also occur in tokamak plasmas, for kinklike\cite{Boozer-PRL} and tearing\cite{Boozer-EPST,Reiman} modes.  It was reported in Ref.~\onlinecite{Boozer-EPST} that external perturbations would be amplified if the equilibrium profile is close to marginal stability for tearing. In Refs.~\onlinecite{Boozer-PRL,Reiman}, it was shown that error fields at the plasma edge are amplified in the core by a factor that is inversely proportional to the marginal stability parameter. Our present result shows that the amplification phenomenon extends also to interchange modes (as also found in a related case in Ref.~\onlinecite{Adler}). We include nonlinear perturbations in our methodology to show that the amplification precipitates an already dormant nonlinear instability.

We use a model system to illustrate the basic phenomena. The calculation is done in slab geometry with an effective gravitational field to model field line curvature. Boundary perturbations are treated as a ripple on the boundary conditions, similar to the Kulsrud-Hahm problem.\cite{Kulsrud-ripple}

\begin{figure}
\includegraphics[width=0.35\textwidth]{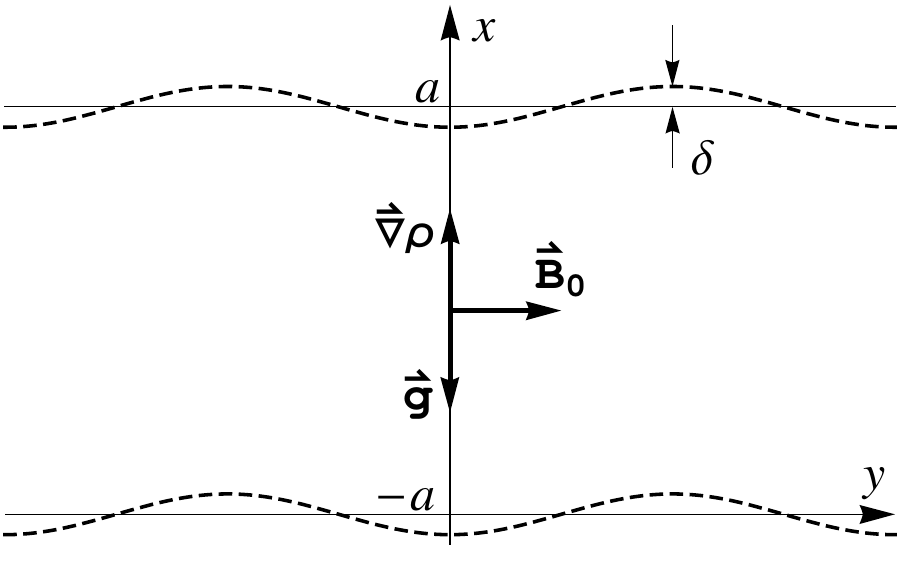}
\caption{\label{fig:system} The conducting plate boundaries, at $x=\pm a$, of a system with $\vec{\mathbf{\nabla}}\rho=\rho_0'\mathbf{\hat{x}}$ opposite a gravitational force $\vec{\mathbf{g}}=-g\mathbf{\hat{x}}$ balanced by a transverse field $\vec{\mathbf{B}}_0=B_0\mathbf{\hat{y}}$ are perturbed by a ripple of amplitude $\delta$.}
\end{figure}

\emph{II. Linear boundary perturbation problem.}---We begin with a simple-minded linear calculation to first demonstrate the amplification phenomenon. Consider an incompressible system describable by the two-dimensional, reduced magnetohydrodynamic (MHD) equations\cite{Strauss} given by,
\begin{equation}\label{eq:flux}
\partial_t\psi=\{\psi,\varphi\},
\end{equation}
\begin{equation}\label{eq:momentum}
\hat{\mathbf{z}}\!\cdot\!\vec{\mathbf{\nabla}}_{\!\perp}\!\!\times\!\rho(\partial_t\vec{\mathbf{u}}+\{\varphi,\vec{\mathbf{u}}\}) = \{\psi,\nabla_{\!\perp}^2\psi\} + g\partial_y \rho,
\end{equation}
\begin{equation}
\vec{\mathbf{u}}=\mathbf{\hat{z}}\!\times\!\mathbf{\nabla}_{\!\perp}\varphi,\quad
\vec{\mathbf{B}}_\perp=\mathbf{\hat{z}}\!\times\!\mathbf{\nabla}_{\!\perp}\psi,
\end{equation}
where, in general, $\rho=\rho(\psi)$ and
\begin{equation}
\{f,h\}\equiv\partial_xf\partial_yh-\partial_yf\partial_xh.
\end{equation}

The problem of interest is illustrated in Fig.~\ref{fig:system} with $\rho_0'$ and $B_0$ constant, and such that we are stable to the ideal MHD interchange mode. The system is taken to be periodic in the $y$ direction, and
\begin{equation}\label{eq:xboundary}
\partial_y\psi=-k\sin(ky)\delta\partial_x\psi
\end{equation}
at $x=\pm a-\delta \cos(ky)$ provides the rippled boundary condition in the $x$ direction. For $\delta=0$ this system is in static equilibrium.

We now perturb the boundaries at $x=\pm a$ by adiabatically introducing $\delta/a \ll 1$ and allowing for a quasistatic equilibrium, $\psi=B_0x+\tilde{\psi}$, to form. In general, $\tilde{\psi}$ satisfies the nonlinear equation
\begin{equation}\label{eq:tildepsieq}
(B_0^2\nabla_\perp^2+g\rho_0')\partial_y\tilde{\psi}=-B_0\{\tilde{\psi},\nabla_\perp^2\tilde{\psi}\},
\end{equation}
and the boundary condition, Eq.~(\ref{eq:xboundary}). The solution
\begin{equation}\label{eq:tildepsisol}
\tilde{\psi}=\frac{\delta B_0}{\cos(k_xa)}\cos(k_xx)\cos(ky),
\end{equation}
where
\begin{equation}\label{eq:grho0prime}
k_x^2=\frac{g\rho_0'}{B_0^2}-k^2,
\end{equation}
satisfies Eq.~(\ref{eq:tildepsieq}) and Eq.~(\ref{eq:xboundary}) to lowest, linear, order in $\delta/a$. From Eq.~(\ref{eq:tildepsisol}), we note that as $k_xa$ approaches $\pi/2$, the perturbation gets amplified. In fact, as we will confirm later, $k_x a = \pi/2$ corresponds precisely to the linear stability criterion for the ideal interchange mode, where the critical transverse field for marginal stability, $B_0=B_c$, is obtained from Eq.~(\ref{eq:grho0prime}) at $k_x=k_c\equiv\pi/2a$.

We can determine the scaling of this amplification by letting $B_0=B_c+b$ and $k_x=k_c-\Delta k_x$, where $b/B_c\sim\Delta k_x/k_c\ll1$. To lowest order in $b/B_c$
\begin{equation}
\Delta k_x = \frac{b}{B_c}\frac{k_\perp^2}{k_c},
\end{equation}
where $k_\perp^2\equiv k_c^2+k^2$, and Eq.~(\ref{eq:tildepsisol}) yields
\begin{equation}\label{eq:tildepsiapprox}
\tilde{\psi}\approx\frac{\delta/a}{b/B_c}\frac{k_c}{k_\perp^2}B_c\cos(k_xx)\cos(ky).
\end{equation}
Therefore, as we approach criticality by letting $b\rightarrow 0$, a small perturbation at the boundary of $\mathcal{O}(\delta/a)$ can induce a large response in the bulk of the plasma, scaling like $B_c\delta/ba$. At exactly $B_0=B_c$ the solution becomes ill-defined away from the boundary.

The solution given by Eq.~(\ref{eq:tildepsisol}) is valid as long as $\delta/a$ is small, such that $B_c\delta \ll ba$. However, this solution clearly establishes the phenomenon of amplification. We show in what follows that nonlinear effects arise at even smaller boundary amplitudes, significantly modifying our understanding of the effect of boundary perturbations on stability.

\emph{III. Nonlinear evolution.}---For a system that is marginally stable to the ideal MHD interchange mode, it was previously shown that small homogeneous perturbations in the plasma can result in nonlinear, explosive growth.\cite{Bagaipo} The nonlinear instability result showed that the optimal scaling of magnetic perturbations was given as  $|\tilde{\psi}/\psi_0|\sim\epsilon\equiv(b/B_c)^{1/2}$. This realization prompts us to apply this nonlinear stability scaling as optimal ordering for nonlinearities in the present amplification calculation. Thus, optimally, we should have $\delta/a\sim\epsilon^3$, using Eq.~(\ref{eq:tildepsiapprox}). With this scaling, we will show that $\delta$ on the boundary can introduce a nonlinear instability in the plasma. It is important to note that even though we order the parameters as described, the boundary perturbation amplitude, $\delta/a$, and marginality condition, $b/B_c$, are independent, small parameters.

Using the marginality condition as a smallness parameter we expand $\psi$ in a series, i.e. let
\begin{equation}
\psi=\psi_0+\psi_1+\psi_2+\psi_3+\cdots,
\end{equation}
where each successive term is smaller by a factor of $\epsilon$ and $\psi_0=(B_c+b)x$. By matching terms order by order, we solve Eq.~(\ref{eq:tildepsieq}), where $\tilde{\psi}=\psi-\psi_0$, using the boundary condition given by Eq.~(\ref{eq:xboundary}).


To order $\epsilon$, Eq.~(\ref{eq:tildepsieq}) yields
\begin{equation}\label{eq:psi1eq}
B_c^2(\nabla_{\!\perp}^2+k_\perp^2)\partial_y\psi_1=0,
\end{equation}
where we have substituted for $g\rho_0'$ using Eq.~(\ref{eq:grho0prime}) with $B_0=B_c$ and $k_x=k_c$. Taking Eq.~(\ref{eq:xboundary}) to lowest order implies that
\begin{equation}
\left. \partial_y\psi_1\right|_{x=\pm a} = 0.
\end{equation}
Including the lowest order term in the solution given by Eq.~(\ref{eq:tildepsisol}), we find that
\begin{equation}\label{eq:psi1sol}
\psi_1=\left[A+\frac{\delta/a}{b/B_c}\frac{k_c}{k_\perp^2}B_c\right]\cos(k_cx)\cos(ky),
\end{equation}
where, for convenience, $A$ is introduced as a free parameter in this particular manner to represent the plasma response. We are interested in how $\delta$ in the boundary induces $A$ in the plasma. We use the scaling $k_cA/B_c\sim\epsilon$ for the plasma perturbation, but it will be treated as a separate small parameter.

The lowest order equation, given by Eq.~(\ref{eq:psi1eq}), can also be arrived at by solving the linear, ideal MHD, interchange mode problem and insisting that $\omega=0$. This results in a zero frequency state, given by Eq.~(\ref{eq:psi1sol}) with $\delta=0$, where Alfv\'{e}nic restoring forces exactly balances the Rayleigh-Taylor growth rate, i.e. $k_\perp^2V_{Ac}^2 - g\rho_0'/\rho_0=0$. From this, we can conclude that $B_c$ is the critical field strength needed to be at marginal conditions. Allowing for $b>0$ means that the system is marginally stable to the ideal MHD interchange mode.

Continuing to order $\epsilon^2$, we find that $\psi_2$ satisfies the equation
\begin{align}
B_c^2(\nabla_{\!\perp}^2+k_\perp^2)\partial_y\psi_2&=-B_c\{\psi_1,\nabla_{\!\perp}^2\psi_1\}\\&=0\label{eq:psi2eq}
\end{align}
and the boundary condition
\begin{equation}
\left. \partial_y\psi_2\right|_{x=\pm a}=0.
\end{equation}
This implies that $\psi_2$ is only a function of $x$. We can solve for $\psi_2$ by taking Eq.~(\ref{eq:flux}) to second order and averaging over $y$ to get
\begin{equation}
\partial_t\psi_2 = \overline{\partial_x(\partial_y\tilde{\varphi}\psi_1)},
\end{equation}
where the bar denotes an average over $y$. Equation (\ref{eq:flux}) to first order implies that
\begin{equation}
\partial_t\psi_1=B_c\partial_y\tilde{\varphi},
\end{equation}
and so we find that
\begin{equation}\label{eq:psi2sol}
\psi_2=-\frac{1}{4}\frac{k_c}{B_c}\left[A+\frac{\delta/a}{b/B_c}\frac{k_c}{k_\perp^2}B_c\right]^2\sin(2k_cx).
\end{equation}
This term represents the flattening of the perturbed field lines (“zonal field”) to second order, driven by the first order perturbation.

In order to find how $\delta$ drives the amplitude $A$ we extend our analysis to order $\epsilon^3$ where Eq.~(\ref{eq:tildepsieq}) yields
\begin{align}\label{eq:psi3eq}
&B_c^2(\nabla_\perp^2+k_\perp^2)\partial_y\psi_3+2B_cb\nabla_\perp^2\partial_y\psi_1=\notag\\&\qquad -B_c(\{\psi_1,\nabla_\perp^2\psi_2\}+\{\psi_2,\nabla_\perp^2\psi_1\}),
\end{align}
with the boundary condition
\begin{equation}
\left.\partial_y\psi_3\right|_{x=\pm a}=-\delta B_ck\sin(ky).
\end{equation}
The above equations imply that $\psi_3=\psi_3(x)\cos(ky)$ where $\psi_3(x)$ is a linear combination of $k_c$ and $3k_c$ harmonic terms. The $3k_c$ harmonic is straightforward and uninteresting; however, the $k_c$ harmonic is secular, so we will focus on obviating this secularity. Using Eq.~(\ref{eq:tildepsisol}) to order $\epsilon^3$ we let
\begin{equation}\label{eq:psi3sol}
\psi_3=B_c\frac{\delta}{a}x\sin(k_cx)\cos(ky).
\end{equation}
Substituting Eqs.~(\ref{eq:psi1sol}), (\ref{eq:psi2sol}), and (\ref{eq:psi3sol}) into Eq.~(\ref{eq:psi3eq}) and insisting that the secular terms go to zero yields
\begin{equation}\label{eq:result}
-2B_cbA+\frac{k_c^2}{4}\frac{k^2-3k_c^2}{k_\perp^2}\left[A+\frac{\delta/a}{b/B_c}\frac{k_c}{k_\perp^2}B_c\right]^3=0,
\end{equation}
after simplification. The above result gives the sought-after relationship between the boundary perturbation amplitude and the amplitude of the plasma response.

To discuss this result we now consider allowing for the plasma response to evolve in time, at a rate slower than Alfv\'{e}n time, $\tau_A=(k_cV_{Ac})^{-1}$, as we distort the boundary at an even slower rate. Explicitly, we let $A=A(t)$ with $\tau_A\partial_t\sim\epsilon$ but we keep $\dot{\delta}$ small. It is easy to show that doing this results in the nonlinear time evolution of $A(t)$ given by
\begin{equation}\label{eq:timeevolution}
\frac{1}{k^2}\ddot{A} = -2bA +\frac{1}{4}\frac{k^2-3}{k^2+1}\left[A+\frac{2}{\pi}\frac{\delta}{b}\frac{1}{k^2+1}\right]^3,
\end{equation}
where we normalize the variables by setting $k_c$, $B_c$, and $V_{Ac}$ equal to 1, for simplicity.
Multiplying Eq.~(\ref{eq:timeevolution}) by $\dot{A}$ and integrating once yields the ``energy'' integral $E_0=\dot{A}^2/2k^2+U(A;\delta)$, where
\begin{equation}\label{eq:potential}
U(A;\delta)=bA^2-\frac{1}{16}\frac{k^2-3}{k^2+1}\left[A+\frac{2}{\pi}\frac{\delta}{b}\frac{1}{k^2+1}\right]^4.
\end{equation}
The above equation represents a potential energy as a function of $A$, given $k$, $b$ and $\delta$, and determines the overall stability of the system. We have chosen $b>0$, to be marginally stable, and so, for $k^2<3$ the system is stable to all perturbations. However, for $k^2>3$ and fixed $b$ the stability of the system is dependent on the size of $\delta$ and $A_0=A(t=0)$.

\begin{figure}
\includegraphics[width=0.35\textwidth]{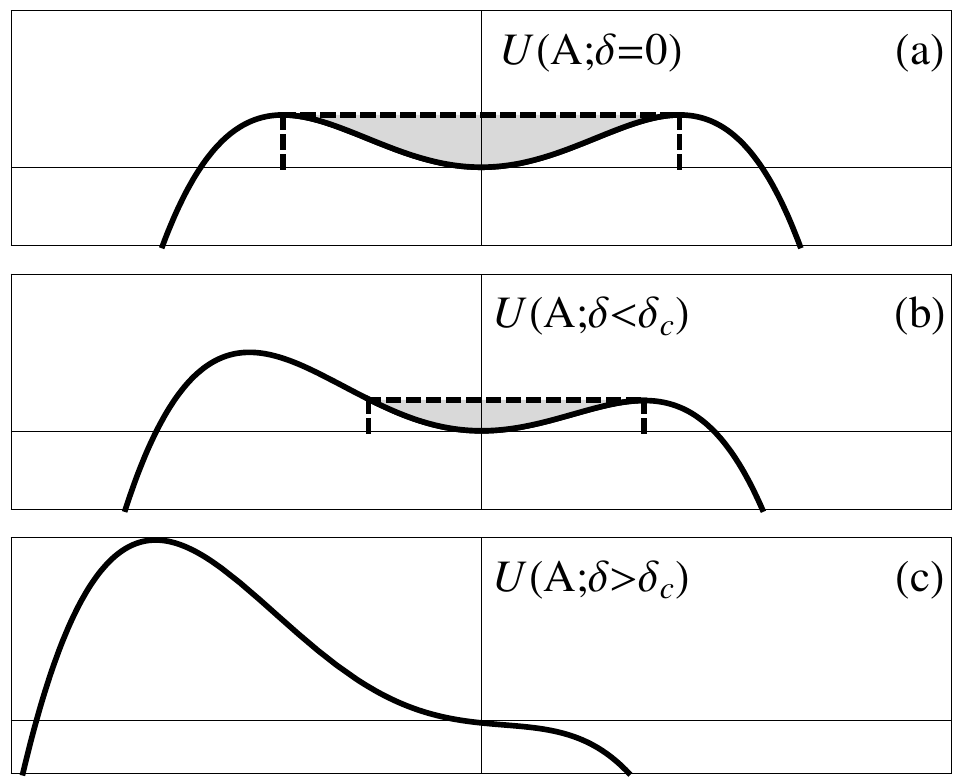}
\caption{\label{fig:potential} Plot of the potential energy $U(A;\delta)$ as a function of amplitude, $A$, for: (a) $\delta=0$; (b) $\delta<\delta_c$; and (c) $\delta>\delta_c$. The dotted box shows the shrinking boundaries of the stable well (shaded region).}
\end{figure}

The potential, with $k^2>3$ and fixed $b$, is shown in Fig.~\ref{fig:potential} for different values of $\delta$. For $\delta=0$ (Fig.~\ref{fig:potential}a), there is a stable well for $|A_0|<A_c$, where
\begin{equation}\label{eq:Ac}
A_c=2\sqrt{2}\sqrt{\frac{k^2+1}{k^2-3}}\!\times b^{1/2}.
\end{equation}
This result is the same as the one found in Ref.~\onlinecite{Bagaipo} where nonlinear stability was noted and simulated for large enough $A$. As $\delta$ is increased the stable well shrinks as the two positive roots of $U'(A;\delta)$ merge. From Fig.~\ref{fig:potential}b we can see that the symmetry of the potential is broken and one side of the stable well drops so that the critical $A_0$ to stay nonlinearly stable is less than the one given in Eq.~(\ref{eq:Ac}). When $\delta=\delta_c$, where
\begin{equation}\label{eq:deltac}
\delta_c = \frac{\pi}{2}\sqrt{\frac{32}{27}\frac{(k^2+1)^3}{k^2-3}}\!\times b^{3/2},
\end{equation}
the positive roots of $U'(A;\delta)$ become degenerate and the stable well becomes a point. When $\delta>\delta_c$ (Fig.~\ref{fig:potential}c) the system is always unstable for any $A_0$. This means that, even though the system is linearly stable, a boundary perturbation of order $(b/B_c)^{3/2}$ can precipitate a response in the plasma of much larger amplitude, of order $(b/B_c)^{1/2}$, accompanied by explosive growth.

We remark that, in this calculation, we have assumed the density gradient, $\rho_0'$, to be constant. This is done for simplicity, to illustrate the two phenomena of amplification and nonlinear instability in a transparent manner. We have shown elsewhere that keeping a general $\rho_0(x)$ did not result in new phenomena or mitigate the appearance of the foregoing phenomena, and that $\rho_0'$ constant is a reasonable simplification.

\emph{IV. Conclusion.}---We have shown that for systems operating close to marginal stability for the interchange mode, a small perturbation on the boundary can induce a large response in the core of the plasma. A simple linear analysis shows that a secondary equilibrium, with amplitude inversely proportional to the marginal stability parameter, can exist as boundary perturbations are added adiabatically. This suggests that the system would need to be far enough away from criticality so as to stay well-defined. We accordingly extend this analysis nonlinearly to show that, even if the perturbation were scaled much smaller than the marginal stability parameter, the system can become nonlinearly unstable for boundary perturbations larger than a critical value.  These results have implications in the stability analysis and corresponding design of magnetic confinement devices operating close to marginal conditions.  With $\Delta\beta>0$, a linear stability analysis would show that the system is stable to all perturbations; however, relatively much smaller boundary perturbations, of order $(\Delta\beta/\beta_c)^{3/2}$, can destabilize the system nonlinearly, with subsequent perturbations growing without small amplitude saturation.

\end{document}